\documentclass[twocolumn,aps,prl,showpacs]{revtex4}
\usepackage[T1]{fontenc}
\usepackage[latin9]{inputenc}
\usepackage{amsmath}
\usepackage{amssymb}
\usepackage{graphicx}
\usepackage{esint}

\makeatletter
\@ifundefined{textcolor}{}
{%
 \definecolor{BLACK}{gray}{0}
 \definecolor{WHITE}{gray}{1}
 \definecolor{RED}{rgb}{1,0,0}
 \definecolor{GREEN}{rgb}{0,1,0}
 \definecolor{BLUE}{rgb}{0,0,1}
 \definecolor{CYAN}{cmyk}{1,0,0,0}
 \definecolor{MAGENTA}{cmyk}{0,1,0,0}
 \definecolor{YELLOW}{cmyk}{0,0,1,0}
}

\makeatother

\begin{document}

\title{Radio-Frequency Manipulation of Fano-Feshbach Resonances in an Ultracold Fermi Gas of $^{40}$K}



\author{Lianghui Huang$^{1}$, Pengjun Wang$^{1}$, B. P. Ruzic$^{2}$, Zhengkun Fu$^{1}$, Zengming
Meng$^{1}$, Peng Peng$^{1}$, J. L. Bohn$^{2}$ and Jing
Zhang$^{1,3\dagger}$}

\affiliation{$^{1}$State Key Laboratory of Quantum Optics and Quantum Optics
Devices, Institute of Opto-Electronics, Shanxi University, Taiyuan
030006, P. R. China\\
$^{2}$JILA, University of Colorado and National Institute of
Standards and Technology, Boulder, Colorado 80309-0440, USA\\
$^{3}$Synergetic Innovation Center of Quantum Information and
Quantum Physics, University of Science and Technology of China,
Hefei, Anhui 230026, P. R. China}

\date{\today}
\begin{abstract}
Experimental control of magnetic Fano-Feshbach resonances in
ultracold $^{40}$K Fermi gases, using radio-frequency (RF) fields,
is demonstrated. Spectroscopic measurements are made of three
molecular levels within 50 MHz of the atomic continuum, along with
their variation with magnetic field. Modifying the scattering
properties by an RF field is shown by measuring the loss profile
versus magnetic field. This work provides the high accuracy
locations of ground molecular states near the s-wave Fano-Feshbach
resonance, which can be used to study the crossover regime from a
Bose-Einstein condensate to a Bardeen-Cooper-Schrieffer superfluid
in presence of an RF field.
\end{abstract}


\maketitle

The capability to tune the strength of elastic interparticle
interactions has led to explosive progress of using ultracold atomic
gases to create and explore many-body quantum systems
\cite{Chin2010}. Magnetic-field-induced Fano-Feshbach resonances are
among the most powerful tools for this purpose, and have been used
widely in atomic gases of alkali atoms. An alternative technique for
tuning interatomic interactions is called optical Feshbach
resonances (OFR) \cite{Fedichev1996,Bohn1999}, in which free atom
pairs are coupled to an electronically excited molecular state by a
laser field tuned near a photoassociation resonance
\cite{Enomoto2008,Yamazaki2010,Blatt2011,Yan2013}. The OFR offers
more flexible control over interaction strength with high spatial
and temporal resolution. Furthermore, laser light in combination
with magnetic Fano-Feshbach resonances have been developed to modify
the interatomic interaction in Bose gases
\cite{Junker2008,Bauer2009,Bauer2009-PRA} and Fermi gases
\cite{Fu2013}.

Radio-frequency (RF) radiation is an appealing alternative means for
manipulating ultracold atoms. Note that for alkali atoms, magnetic
field Fano-Feshbach resonances may be difficult to find, such as in
$^{87}$Rb; and that OFR's are problematic in these species because
of large losses due to spontaneous emission. Manipulation of
scattering lengths via RF radiation therefore represents a powerful
new tool. In the context of ultracold gases, RF can be used to
couple a two-atom scattering state to a bound molecular state
(free-bound coupling) similarly to the OFR. RF also can drive
transitions between bound states (bound-bound coupling). As a probe,
RF has been used extensively to determine the $s$-wave scattering
length near a Feshbach resonance by directly measuring the RF shift
induced by mean-field interactions \cite{Regal2003b}, to demonstrate
many-body effects and quantum unitarity \cite{Gupta2003}, and to
probe the occupied spectral function of single-particle states and
the energy dispersion through Bose-Einstein condensate (BEC) -
Bardeen-Cooper-Schrieffer (BCS) crossover \cite{Stewart2008}. RF can
also be considered as a means of controlling scattering length in a
variety of scenarios. Zhang et al. \cite{Zhang2009} proposed to
independently control different scattering lengths in multicomponent
gases using RF dressing. The RF coupling of magnetic Fano-Feshbach
resonances in a $^{87}$Rb Bose gas has been studied experimentally
and theoretically \cite{Kaufman2009,Hanna2010}. Tscherbul et al.
\cite{Tscherbul2010} performed a theoretical analysis of
manipulating Feshbach resonances of $^{87}$Rb with RF field.
Papoular et al. \cite{Papoular2010} suggested using a microwave
field to control collisions in atom gases at zero magnetic field.
Further, Avdeenkov \cite{Avdeenkov12_PRA} applied this same idea to
manipulate scattering of polar molecules.

In this paper, we experimentally investigate a magnetic
Fano-Feshbach resonance in combination with an RF field in ultracold
$^{40}$K Fermi gases. We measure the spectrum of the nearby
molecular bound states with partial-wave quantum number $L=0$ by
applying a near-resonant RF field. We also measure the free-to-bound
transition from free atoms with attractive interaction to these same
molecular states. For all three states measured, the binding
energies are in good agreement with a theoretical calculation.
Further, the loss of atoms versus magnetic field is measured, to
determine the ability of the RF field to modify scattering. The
position of the narrow loss features induced by the RF field in the
broad loss profile of magnetic Fano-Feshbach resonances can be
changed easily by setting the frequency of RF field, which
represents the modification of the scattering properties near a
magnetic Fano-Feshbach resonance, producing resonance features
narrower in magnetic field than the original resonance.


We consider potassium 40 atoms in a mixture of hyperfine states
$|F,M_F \rangle = |9/2, -9/2\rangle$ and $|9/2,-7/2 \rangle$, where
$F$ and $M_F$ are the total (electronic plus nuclear) spin and its
projection on the magnetic field axis, respectively. Atoms in such a
mixture are known to have an s-wave Fano-Feshbach resonance at a
magnetic field $B=202.2$ G \cite{Chin2010}. We consider interspecies
collisions between these two states, and their nearby molecular
bound states, given in the atom-pair quantum numbers $|F_1, M_{F_1}
\rangle | F_2,M_{F_2} \rangle |L,M_L \rangle$, as described in
Hund's coupling case (e). Here $L$ and $M_L$ are the quantum numbers
of the partial wave angular momentum and its projection. In an
ultracold gas, the atoms start in either free pairs with $L=0$, or
else in a very weakly-bound Feshbach molecular state with $L=0$.
Since we consider only RF transitions that cannot change $L$, we
omit this index in what follows.

\begin{figure}[t]
\begin{centering}
\includegraphics[clip,width=0.4\textwidth]{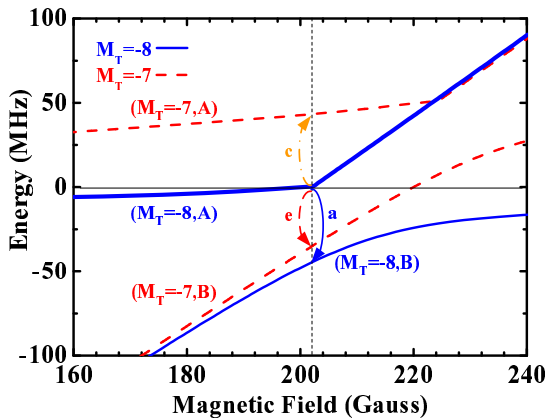}
\par\end{centering}

\caption{(Color online) Energy spectrum of relevant molecular levels
of $^{40}K_{2}$ versus the magnetic field in the electronic ground
state. The zero of energy is taken to be that of two separated atoms
$|9/2,-9/2\rangle  + |9/2,-7/2\rangle$ at each magnetic field
strength. $M_T$ is total angular momentum projection
$M_{F_1}+M_{F_2}$}

\label{fig1}
\end{figure}

Figure 1 shows the nearby bound s-wave ($L=0$) molecular levels
versus magnetic field, as calculated by a coupled channel model.
This model has been engineered to fit simultaneously the s-wave and
p-wave Fano-Feshbach resonances reported in Ref. \cite{Gaebler2010},
and should be a reasonable representation of potassium cold
collisions near $B=200$ Gauss. In this figure the zero of energy
corresponds to the threshold energy of the entrance channel
$|9/2,-9/2\rangle + |9/2,-7/2\rangle$. The figure shows two
molecular bound states with total angular momentum projection $M_T
\equiv M_{F_1}+M_{F_2} = -8$ (blue solid lines), which exhibit an
avoided crossing at the resonance. In the upper curve ($M_T=-8,A$),
the line represents the Feshbach molecule state for $B<B_0$, whereas
it becomes a scattering resonance at $B>B_0$. The lower curve
($M_T=-8,B$) remains a relatively deeply bound (by $\sim 45 $ MHz)
state in the field range shown. These states can be reached by RF
radiation polarized with the magnetic field along the quantization
axis set by the magnetic field, satisfying the selection rule
$\Delta M_T=0$. In addition, the figure shows two bound states with
$M_T=-7$, which can be reached by RF with perpendicular
polarization, with selection rule $\Delta M_T= \pm1$. The upper
($M_T=-7,A$) of these bound states is weakly bound with respect to
the $|9/2,-9/2\rangle + |7/2,-7/2\rangle$ threshold, and becomes
unbound into this continuum at $B \approx 220$ Gauss. The figure
also shows arrows (labeled a, c, e) indicating the allowed
bound-to-bound transitions that are measured in the experiment.

The experimental apparatus has been described in our previous work
\cite{Xiong2008,Xiong2010a,Xiong2010b,Wang2011}. The degenerate
Fermi gas of about $N\simeq2\times10^{6}$ $^{40}$K atoms in the
$|9/2,9/2\rangle$ internal state is obtained with $T/T_{F}\simeq0.3$
by evaporatively sympathetic cooling with bosonic $^{87}$Rb atoms in
the $|2,2\rangle$ state inside a crossed optical trap. Here $T$ is
the temperature and $T_{F}$ is the Fermi temperature defined by
$T_{F}=E_{F}/k_{B}=(6N)^{1/3}\hbar\overline{\omega}/k_{B}$ with a
geometric mean trapping frequency $\overline{\omega}$. Then Fermi
gas with equal spin-population in the $|9/2,-9/2\rangle$ and
$|9/2,-7/2\rangle$ states is prepared at about $B\simeq219.4$ G.
These two hyperfine states form the incoming state $|9/2,-9/2\rangle
+ |9/2,-7/2\rangle$ in the entrance channel for a pair of atoms, as
shown in Fig. \ref{fig1}. We use a magnetically controlled
Fano-Feshbach resonance at $B_{0}=202.20\pm0.02$ G to adiabatically
convert a pair of atoms into extremely weakly bound molecules
(binding energy $< 100$ kHz).

We place the coil just outside the glass cell and the oscillating
magnetic field generated by the RF coil may be parallel (or
perpendicular by changing the positions of the RF coil) to the
magnetic bias field of the Fano-Feshbach resonance. We apply the RF
pulse in a rectangular temporal shape, with variable time that
depends on the measurement being made. Near resonance with one of
the bound-to-bound transitions, the RF field induces loss in the
population of Feshbach molecules due to the excitation to the other
molecular states. In order to determine the number of remaining
Feshbach molecules in the trap, after turning off the RF, another
gaussian-shape RF pulse with duration about 400 $\mu s$ is applied
to dissociate the remaining molecules into free atoms in the state
$|9/2,-9/2\rangle\ + |9/2,-5/2\rangle$. The frequency of the RF
pulse used to dissociate Feshbach molecules is fixed to a value that
is about $E_{b}$ kHz larger than the Zeeman splitting between the
hyperfine states $|9/2,-7/2\rangle$ and $|9/2,-5/2\rangle$ for the
certain magnetic field, which corresponds to the transition from the
bound molecules to the free atom state $|9/2,-9/2\rangle\ +
|9/2,-5/2\rangle$. After the dissociation RF pulse, we abruptly turn
off the optical trap and magnetic field, and let the atoms
ballistically expand for $12$ ms in a magnetic field gradient
applied along the $\hat{y}$ axis and then take absorption image
along the $\hat{z}$ direction. The atoms in different hyperfine
states $N_{\sigma}$ ($\sigma=|-7/2\rangle,|-5/2\rangle...$) are
spatially separated and analyzed, from which we determine the
fraction $N_{-5/2}/(N_{-5/2}+N_{-7/2})$ for different RF frequencies
to obtain the the spectrum of the bound molecular states.

\begin{figure}[t]
\begin{centering}
\includegraphics[clip,width=0.45\textwidth]{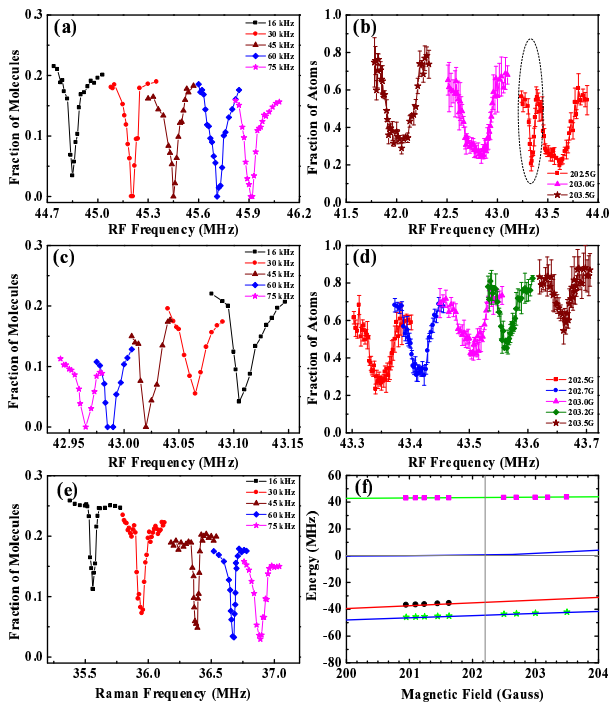}
\par\end{centering}

\caption{(Color online) Bound-to-bound and free-to-bound
spectroscopy of ground $^{40}K_{2}$ molecules for different magnetic
field. (a) The bound-to-bound spectroscopy from Feshbach molecular
state to deeply bound molecular state ($M_T=-8,B$). (b) The
free-to-bound spectroscopy from free atom to deeply bound molecular
state ($M_T=-8,B$). (c) The bound-to-bound spectroscopy from
Feshbach molecular state to deeply bound molecular state
($M_T=-7,A$). (d) The free-to-bound spectroscopy from free atom to
deeply bound molecular state ($M_T=-7,A$). (e) The bound-to-bound
spectroscopy of deeply bound molecular state ($M_T=-7,B$) for
different magnetic field, which is obtained by two-color stimulated
Raman process. (f) The energies of deeply bound molecular state as
the function of magnetic field are obtained from (a)-(e). The dot
points are the experimental data and the solid lines are obtained
from theoretical calculation.}

\label{fig2}
\end{figure}

Figure \ref{fig2} reports the results of the spectroscopic
measurements for the transitions described in Fig. \ref{fig1}.
Figure \ref{fig2}(a) shows the bound-to-bound transition (labeled as
"a" in Fig. \ref{fig1}) near 45 MHz for different magnetic fields
corresponding to different binding energies $E_{b}$ of the Feshbach
molecules. Here the bound molecules are illuminated with the RF
pulse duration time of $5$ ms and the RF field is parallel to the
direction of the magnetic bias field. The lifetimes of deeply bound
molecular states are measured to be less than 2 ms, which are much
shorter than the lifetimes of Feshbach molecules. When the RF field
is perpendicular to the direction of the magnetic bias field, we do
not observe any loss of the Feshbach molecules, confirming that the
bound state is a state of $M_{T}=-8$. It is also possible to measure
this state starting from a pair of free atoms at $B>B_0$, as shown
in Fig. \ref{fig2}(b). Here the free atoms are illuminated with the
RF pulse duration time of $50$ ms. From these data, we reproduce the
binding energy of this state versus magnetic field, which is plotted
as the lower data set in Fig. \ref{fig2}(f), and compared directly
to the theoretical calculation (blue line).

Figure \ref{fig2}(c) shows the bound-to-bound transition (labeled as
"c" in Fig. \ref{fig1}) near 43 MHz for different magnetic fields,
corresponding to the transition from Feshbach molecular state to the
upper branch molecular state ($M_T=-7,A$). Again this transition is
identified by the energetics of the transition and the slope of
transition energy with respect to magnetic field, as computed in the
model. The transition can be driven, as expected, by RF radiation of
perpendicular polarization, exploiting the selection rule $\Delta
M_T=-1$. We also note, to our surprise, that resonant features
appear at the same transition frequencies for parallel polarization,
with selection rule $\Delta M_T=0$. The model is unable to identify
such a state in the spectrum, and the appearance of these lines in
parallel polarization remains a mystery. This state can also be
identified in  free-to-bound spectroscopy for magnetic fields above
$B_{0}=202.2$ G, Fig. \ref{fig2}(d). Comparing with the theoretical
calculation, the measured bound-to-bound (free-to-bound) transition
corresponds to the transitions near 43 MHz from Feshbach molecular
state (free atoms) to the upper branch molecular state ($M_T=-7,A$)
as shown in the upper data set in Fig. \ref{fig2}(f). Note that the
resonant position (43.35 MHz) of the free-to-bound spectra for the
magnetic bias field 202.5 G in Fig. 2(d) corresponds to the narrow
dip in Fig. 2(b) with the same magnetic bias field, since both
transitions are nearly degenerate at this field.


We have also identified the lower branch molecular state
($M_T=-7,B$). However, we can not observe this state via loss of the
Feshbach molecules, even when applying the maximum power of RF field
in our experimental setup, either in  perpendicular or parallel
polarization. This is presumably because of a comparatively small
Franck-Condon overlap between the states. However, one can use a
pair of laser beams to coherently couple two bound molecular states
via a common electronically excited molecular state (two-color
stimulated Raman process) to enhance the coupling between two ground
bound-bound molecular states. Here, we carefully choose the one
photon detuning of the Raman lasers (wavelength 772.4 nm) to avoid
loss due to a Feshbach resonance induced by the laser between the
ground Feshbach molecular state and the electronically excited
molecular state \cite{Fu2013,Fu2013a}. The configuration of the two
Raman lasers has been described in our previous work \cite{Fu2014}.
Using these Raman lasers we measure bound-to-bound spectroscopy for
ground $^{40}K_{2}$ molecules near 36 MHz for different magnetic
fields, as shown in Fig. 2(e). These binding energies are again in
good agreement with the theoretical model, as shown by the red line
in Fig. \ref{fig2}(f).

\begin{figure}[t]
\begin{centering}
\includegraphics[clip,width=0.35\textwidth]{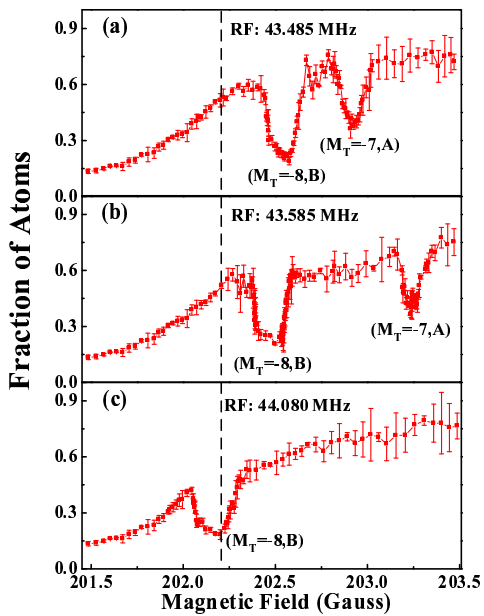}
\par\end{centering}

\caption{(Color online) Atom loss from the entrance channel
$|9/2,-9/2\rangle,|9/2,-7/2\rangle$ in the vicinity of the Feshbach
resonance 202.2 G for different frequencies of RF field. (a) The
frequency of RF field is 43.485 MHz. (b) 43.585 MHz. (c) 44.080
MHz.}

\label{fig3}
\end{figure}

We now consider how the Fano-Feshbach resonance is modified in the
presence of the RF field. We do this by measuring the loss profile
versus magnetic field, for various RF couplings. Two spin mixture
states $|9/2,-9/2\rangle$ and $|9/2,-7/2\rangle$ are initially
prepared at the magnetic field of 210 G and then the field is ramped
quickly to its final value within 3 ms. Then the atoms are
illuminated with the RF pulse for $50$ ms with the RF field
polarized parallel to the direction of the magnetic bias field.
After this, the fraction of Feshbach molecules remaining is counted
as above.

The resulting loss profiles as functions of the magnetic field for
the different RF frequencies are shown in Fig. 3. The broad $s$-wave
Fano-Feshbach resonance of two spin mixture states
$|9/2,-9/2\rangle$ and $|9/2,-7/2\rangle$ is at $B_{0}=202.2$ G
(vertical dashed line in the figure)with a width of 7.04 G
\cite{Gaebler2010}. The maximum atom loss is not centered on this
resonance, but rather occurs at lower-field regions of the spectrum
(the BEC side of resonance), regardless of the RF frequency.  Thus
the main loss occurs where the Feshbach molecular state is already
quite deeply bound \cite{Dieckmann2002,Bourdel2003,Regal2004} in
sharp contrast with the bosonic case, where maximum loss is observed
primarily on the resonance \cite{Wieman2000,Marte2002,Grimm2003}.
In addition, there are narrow loss features in the broad loss
profile when the RF field is applied, which we attribute to
transitions "a" and "c" in Figure \ref{fig1}, namely, transitions to
the bound states ($M_T=-8,B$) and ($M_T=-7,A$). The former moves to
lower magnetic fields as the RF frequency is increased, while the
latter moves to higher magnetic field, as described by the energies
of these resonances. At an RF frequency of $44.080$ MHz, the "a"
transition coincides in magnetic field with the initial resonance,
at $B=202.2$ G. In this case the naturally occurring magnetic
Fano-Feshbach resonance and the RF-coupled bound state can interfere
with one another. The profile of the scattering length can be
deduced from the loss profile. Thus this shows that RF radiation may
be used to place narrow resonances at any desired magnetic field
(which means to locate on a desired position on a broad magnetic
Fano-Feshbach resonance), opening new prospects for control of
collisions. Moreover, our experimental scheme may be modified with
the two optical fields coupling two ground molecular states through
an excited molecular state. The dark molecular state generated by
this protocol can be used to control the interaction strength near
an magnetic Feshbach resonance, and suppress spontaneous emission by
quantum interference \cite{Wu2012}.


In conclusion, we have experimentally demonstrated the technology of
tuning a magnetic-field Fano-Feshbach resonance using an RF field in
ultracold atomic Fermi gases. The spectrum of the nearby molecular
states with partial-wave quantum number $L=0$ is measured by
applying an RF field near the s-wave Fano-Feshbach resonance. By
comparing to a theoretical calculation, we have identified the
quantum numbers of the molecular states. We have moreover
characterized the loss profile of the Fano-Feshbach resonance in the
presence of the RF field. Since RF radiation is easily manipulated,
this technology could be used to switch scattering lengths rapidly
and precisely. The tunability of interatomic interactions, as
demonstrated in this work, provides a new way to explore the
fascinating quantum many-body system of strongly interacting Fermi
gases.

\begin{acknowledgments}
This research is supported by the National Basic Research Program of
China (Grant No. 2011CB921601), NSFC (Grant No. 11234008,
11361161002), NSFC Project for Excellent Research Team (Grant No.
61121064), and Doctoral Program Foundation of the Ministry of
Education China (Grant No. 20111401130001). BPR and JLB acknowledge
funding from an AFOSR MURI grant.

$^{\dagger}$Correspondence should be addressed to Jing Zhang
(jzhang74@aliyun.com, jzhang74@sxu.edu.cn).\end{acknowledgments}

\end{document}